# Mechanism of Anisotropic Crystallization and Phase Transitions under Van der Waals Squeezing

Yuxiang Gao[1,2,3], Zhicheng Zhong[1,2,4,*]
[1]School of Artificial Intelligence and Data Science, University of Science and Technology of China, Hefei 230026, China
[2]Suzhou Institute for Advanced Research, University of Science and Technology of China, Suzhou 215123, China
[3]Suzhou Big Data & AI Research and Engineering Center, Suzhou 215123, China
[4]Suzhou Lab, Suzhou 215123, China

**ABSTRACT** Mechanical confinement strategies, such as van der Waals (vdW) squeezing, have emerged as promising routes for synthesizing non-vdW two-dimensional (2D) layers, surprisingly yielding high-quality single crystals with lateral sizes approaching ~100 μm. However, the underlying mechanisms by which such a straightforward approach overcomes the long-standing synthesis challenges of non-vdW 2D materials remains a puzzle. Here, we investigate the crystallization dynamics and phase evolution of Bi under vdW confinement through molecular dynamics (MD) simulations powered by a machine-learning force filed fine-tuned and distilled from a pre-trained model with DFT-level accuracy. We reveal that pressure-dependent layer modulation arises from a quantum confinement–driven anisotropic crystallization mechanism, in which out-of-plane layering occurs nearly two orders of magnitude faster than in-plane ordering. Two critical transitions are identified: an α-to-β phase transformation at 1.64 GPa, and a subsequent collapse into a single-atomic layer at 2.19 GPa. The formation of large-area single crystals is enabled by substrate-induced orientational selection and accelerated grain boundary migration, driven by atomic diffusion at elevated temperatures. These findings resolve the mechanistic origin of high-quality 2D crystal growth under confinement and establish guiding principles for the controlled synthesis of metastable 2D single crystals, with implications for next-generation quantum and nanoelectronic devices.

*Introduction*—2D materials represent a paradigm shift in quantum material design, hosting unconventional phenomena absent in their three-dimensional (3D) counterparts, such as symmetry-protected topological transport, dimensionally enhanced superconductivity and single-element ferroelectricity. [1–5] However, traditional synthesis approaches like molecular beam epitaxy and chemical vapor deposition are predominantly effective for vdW materials with quasi 2D structures. [6,7] For non-vdW materials with intrinsic thermodynamic instability at atomic thin limit, these approaches typically yield polycrystalline films with sub-micrometer domains or strong substrate bonding. [8–10] This structural degradation severely compromises the intrinsic electronic and transport properties of non-vdW 2D systems, thus highlights the urgent need for alternative fabrication strategies that circumvent these thermodynamic constraints.

Recent breakthroughs in mechanical confinement strategies have begun addressing the synthesis challenge of non-vdW 2D materials. [11–13] Notably, Zhang et al. developed a vdW squeezing method with broad potential for fabricating non-vdW 2D crystals by pressing a thick metal in liquid forms at an elevated temperature for thickness reduction. [12] The achieved 2D metals are single-crystal domains up to 100 μm, representing a two-order-of-magnitude improvement in lateral dimensionality. Despite these advances, critical questions remain unresolved regarding the atomic-scale mechanisms enabling the growth of high-quality 2D single crystals which are intrinsically unstable and pressure-mediated 3D-to-2D phase transitions. The absence of mechanistic understanding severely limits predictive control over layer thickness uniformity and metastable phase selection.

Here, using MD simulations powered by a machine-learning force filed fine-tuned and distilled from a pre-trained model with DFT-level accuracy, we investigated the crystallization and phase transitions of Bi under vdW squeezing. Our simulations reproduced the experimentally observed pressure-dependent layer modulation and predicted two critical transitions: an α-to-β phase transformation at 1.64 GPa, and the formation of strictly single-atomic layers at 2.19 GPa. A quantum confinement–driven anisotropic mechanism was revealed, in which out-of-plane layering not only precedes in-plane ordering, but also occurs at a significantly faster rate. The formation of high-quality single crystals was achieved by substrate-guided orientational selection and promoted atomic diffusion at elevated temperatures. These findings not only elucidate by which mechanical confinement enables crystallization in thermodynamically unstable systems, but also provide fundamental insights for the controlled synthesis of 2D single crystals essential for future nanoelectronic and quantum technologies.

*Methods*—In this work, MD simulations were performed using Large-scale Atomic/Molecular Massively Parallel Simulator (LAMMPS) software [14] with a well-trained deep potential (DP) model. The DP model was generated using the PFD-kit package [15], yielding root-mean-square errors of 6.47 meV/atom in energy and 0.84 eV/Å in atomic forces across a range of $MoS_2$–Bi–$MoS_2$ heterostructures. Unlike conventional ML force fields that often struggle to achieve such accuracy in complex heterostructures, our model achieves DFT-level accuracy owing to its fine-tuning and distillation from a universal pretrained potential. More details of the training dataset, model training and validation are provided in supplementary section I.

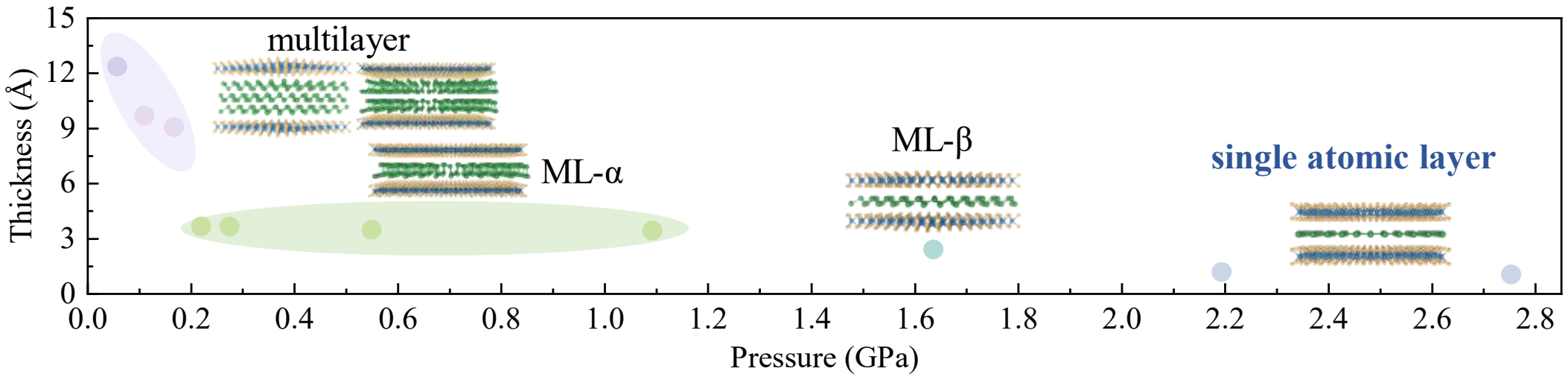

FIG. 1. Pressure-dependent thickness modulation by vdW squeezing.

*Results*—Following the experimental protocol of Zhang et al. [12], we performed MD simulations of Bi under vdW squeezing at 280 °C across a range of applied pressures. Experimental setups are typically limited to pressures below 200 MPa, leaving it unclear whether higher pressures could further reduce Bi thickness to the single-atomic-layer limit. To explore this, we extended simulations up to 2.75 GPa, well beyond the experimental range. The resulting structures exhibit pressure-dependent layering, transitioning from multilayers to strictly single-atomic layers (Fig. 1). A clear trend of decreasing Bi thickness with increasing pressure is observed, consistent with experiments. The monolayer α-phase (ML-α) emerges at 0.22 GPa and remains stable up to 1.09 GPa. At 1.64 GPa, a 1 Å thickness reduction signals a transition to the buckled honeycomb β-phase (ML-β). Further compression to 2.19 GPa yields a strictly single-atomic layer (~1 Å thick). These results demonstrate that pressure serves as a key thermodynamic parameter for accessing symmetry-breaking phases beyond the reach of conventional synthesis.

The ML-α and ML-β phases retain long-range in-plane order after cooling to 298 K and pressure release, indicating that $MoS_2$ encapsulation effectively stabilizes ML Bi (Fig. S2a and S2b). In contrast, the single-atomic layer, initially disordered at 280 °C, crystallizes into nanocrystalline square lattice upon cooling (Fig. S2c). Notably, this square lattice closely resembles structures observed in previous experimental studies [4,16]. However, the square structure spontaneously transformed into ML-β phase upon pressure release, due to unsaturated out-of-plane dangling bonds. [17] The formation energies of ML and single-atomic layer Bi intercalated in bilayer $MoS_2$ were provided in supplementary section II.

Figure 2a shows the time evolution of the interlayer separation between top and bottom $MoS_2$ layers under vdW squeezing at pressures from 0.22 to 2.75 GPa. The interlayer separation decreases in discrete steps, with three intermediate plateaus at 16.9 Å, 14.9 Å, and 11.7 Å, before stabilizing at ~8.9 Å. As pressure increases, the steps shorten in duration but remain consistent in spacing. These quantized transitions reflect the absence of liquid-like behavior and the sequential stabilization of distinct 2D Bi configurations.

We further analyzed the structural evolution of Bi at these steps. Figures 2b and 2c present the time-dependent out-of-plane atomic density profile and radical distribution function (RDF) of Bi during vdW squeezing at 0.22 GPa. The snapshots at 20 ps, 40 ps, and 65 ps correspond to steps I, II, and III, respectively. At step I, Bi atoms adopt a layered arrangement consisting of approximately five layers, though the second and third layers from the top remain diffuse. At step II, Bi atoms reorganize into four well-defined layers with a uniform interlayer spacing of ~3 Å. During step III, the structure further collapses into three distinct layers. Despite clear out-of-plane ordering in steps II and III, the RDFs resemble the initial liquid state, indicating no long-range in-plane order until ~80 ps, when peaks at 6.7 Å and 7.6 Å mark the onset of crystallization. Notably, the formation of out-of-plane layering occurs within a few picoseconds, approximately two orders of magnitude faster than the that of in-plane order, highlighting a anisotropic crystallization pathway.

Such sequential ordering—out-of-plane followed by in-plane—directly contrasts with 3D systems where anisotropic compression induces simultaneous multi-directional ordering. This contrast underscores the essential role of dimensionality in governing the ordering behavior of quasi-2D systems, where confinement is limited to the out-of-plane direction. Similar confinement-induced layering has been widely reported in colloidal crystals and is well described by the confined hard-sphere model, where the system maximizes its entropy (or average free volume) by adopting an layered density profile [18]. Related

*Contact author: zczhong@ustc.edu.cn

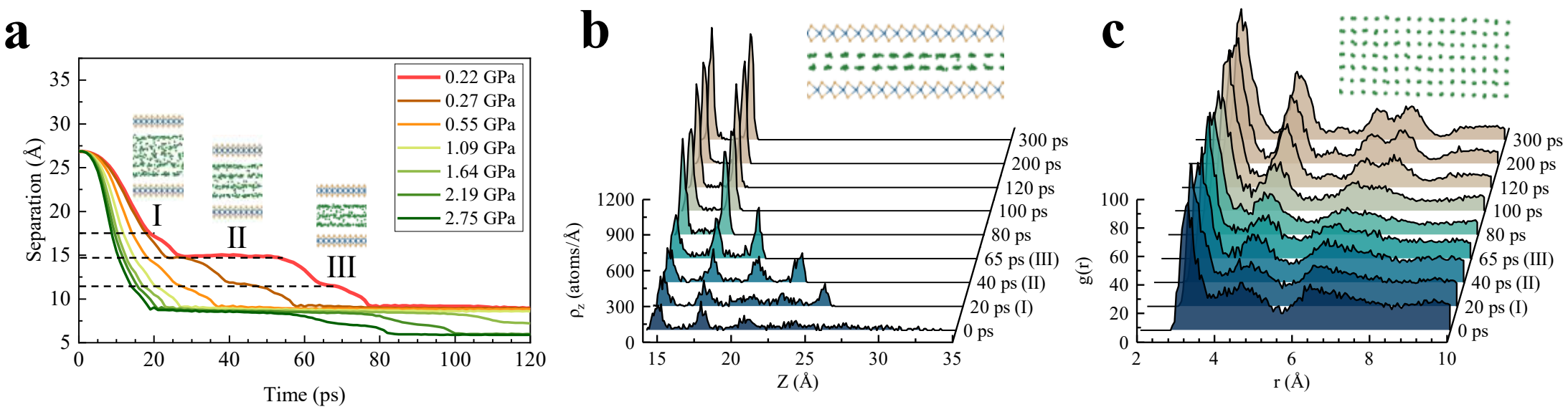

FIG. 2. Multi-step structural transitions of Bi under vdW squeezing. (a) Time evolution of the interlayer separation between the top and bottom $MoS_2$ layers under pressures from 0.22 GPa to 2.75 GPa. Insets show corresponding side views of Bi layers at step I, II, and III. (b) Time evolution of the out-of-plane (z-axis) density profile of Bi layers under 0.22 GPa. Inset show the side view of the Bi layer at 300 ps. (c) Time evolution of the radical distribution function of Bi under 0.22 GPa. Inset shows the top view of the Bi layer at 300 ps.

phenomena have also been observed in confined 2D water, Fe-Ni alloys, and noble gas films. [19–21]

To further elucidate this confinement-driven ordering mechanism, we simulated Bi layers under controlled out-of-plane confinement using virtual top and bottom boundaries. Bi atoms experienced perfectly elastic collisions at the boundaries, restricting motion along the *z*-axis. Figures 3a and 3c display the out-of-plane atomic density profiles and local atomic energies for confinement separations of 4 Å and 2 Å, respectively. Under confinement, the potential energy landscape of Bi develops pronounced energy wells near the boundaries, stabilizing discrete bilayer configurations. As the separation decreases, the depth of these energy wells increases from 0.05 eV at 4 Å to 0.24 eV at 2 Å. This enhanced trapping leads to sharper and more intense density peaks, indicating stronger layering and reduced structural delocalization. Importantly, these pressure- and substrate-free simulations directly demonstrate that out-of-plane ordering arises intrinsically from confinement. This insight not only clarifies the origin of layering in vdW-squeezed Bi, but also provides a general framework for confinement-induced phase behavior in other 2D systems.

Figures 3b and 3d show the corresponding in-plane atomic configurations. At 4 Å separation, Bi atoms spontaneously crystallize into the rectangular (α) phase, while at 2 Å it transforms into the buckled honeycomb (β) phase. Interestingly, this transformation sequence—from honeycomb to rectangular lattice with increasing confinement—reverses the trend observed in colloidal systems and noble gases with isotropic interactions. [22,21] This contrast underscores that in-plane ordering in confined Bi is governed by directional enthalpic interactions, rather than simple steric packing.

However, the simulated monolayers exhibit poor structural integrity with numerous voids, contrasting sharply with experimentally observed micron-scale single crystals. This discrepancy highlights a critical limitation of the rigid-boundary model: the absence of dynamic substrate interactions. In practice, van der Waals substrates like $MoS_2$ can serve as active templates, offering commensurate potential landscapes that promote large-area single-crystal formation. [23] To investigate the influence of substrate interactions, we intercalated a Bi disk with a radius of 40 Å between bilayer $MoS_2$ (Fig. S4a). The θ is defined as the twist angle between the Bi [100] direction and the $MoS_2$ zigzag direction. Fig. 4a illustrates the total energy dependence with θ, exhibiting pronounced minima around 15° and 45°, indicating preferred alignment configurations. These discrete energetic drops are mediated by structure reconstruction (Fig. S4b and S4c). Thus, the $MoS_2$ substrate exerts a strong orientational modulation on the nucleation of the Bi lattice at specific epitaxial angles. These results are consistent with experimental observations that ML Bi aligns with one of the two monolayer $MoS_2$ when twist angle approaches 30°. [12]

To further validate this substrate-guided orientational selection mechanism, we conducted five simulations under 0.22 GPa and 0.27 GPa. Figure 4b summarizes the orientation angles between the Bi lattice and $MoS_2$ at both nucleation and final crystal stages. The nuclei initially exhibit randomly distributed orientation angles, while only those grains that are aligned (or nearly aligned) with the $MoS_2$ lattice continue to grow and eventually dominate. Fig.S5 depicts corresponding MD trajectories, illustrating the transition from a disordered state to a well-ordered Bi monolayer.

Fig. 4c shows the simulated diffraction pattern at early nucleation stage. Diffraction spots of Bi are distinguishable from the six-fold-symmetric spots of $MoS_2$ (marked in dashed orange lines). The Bi spots are relatively diffuse, indicating limited coherence and the lack of long-range order. As growth proceeds, grains with similar orientations coalesce, while misaligned grains dissolve due to unfavorable interfacial energetics. Finally, a single-domain crystal is achieved, evidenced by the sharp and concentrated diffraction pattern (Fig. 4d). Notably, the Bi diffraction spots are perfectly aligned with those of $MoS_2$, confirming that Bi is epitaxially aligned with the $MoS_2$ module. These results provide direct evidence that substrate-induced orientational selection is the key mechanism governing single-crystalline Bi formation, with $MoS_2$ acting as both a stabilizing layer and an epitaxial template, which is consistent with experimental findings. [12]

*Contact author: zczhong@ustc.edu.cn

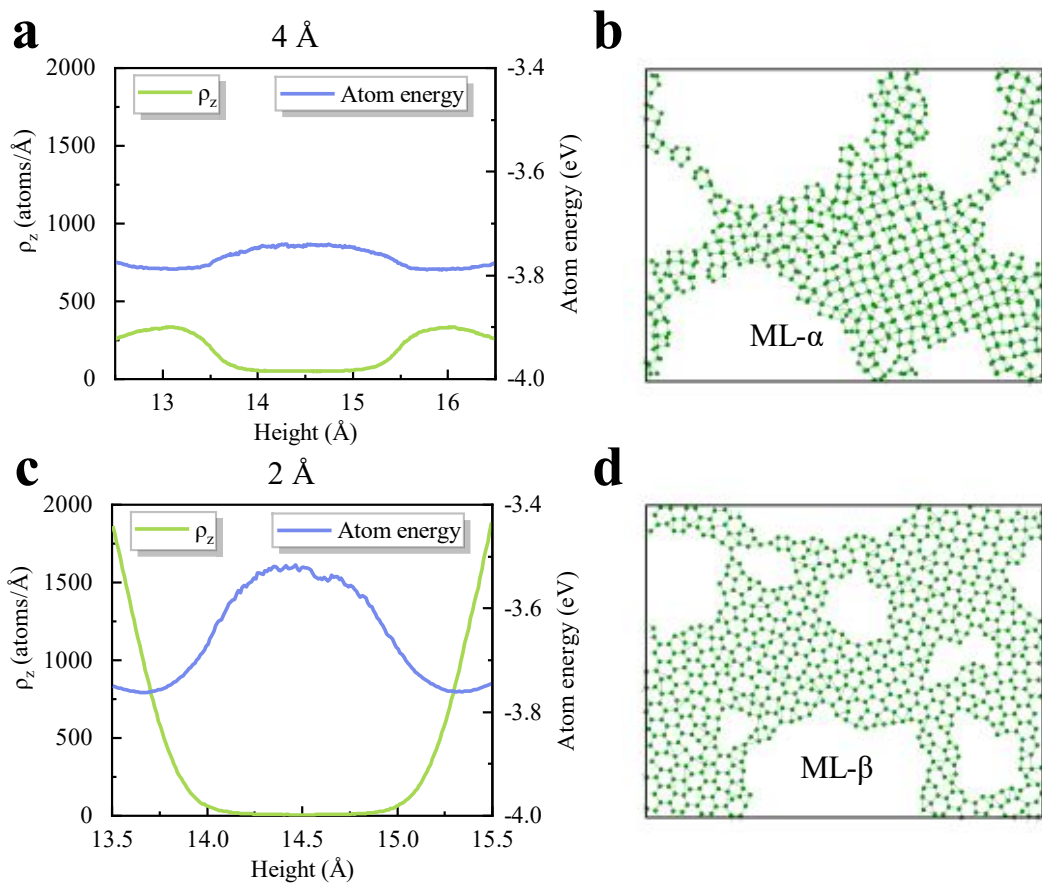

FIG. 3. Spontaneous crystallization and phase transition under confinement. (a and c) The out-of-plan density profile and atom energy of Bi layers with confinement of 4 Å and 2 Å, respectively. (b and d) Corresponding top views.

While energy-driven orientational selection governs the final single-crystal orientation, the formation kinetics is dominated by grain boundary migration and atomic transport. Crucially, the exponential temperature dependence of atomic diffusion profoundly accelerates these processes, as evidenced by our simulations. As shown in Fig. S6, atomic configurations after 50 ps of vdW squeezing at 553–900 K reveal increasing atomic displacements near grain boundaries with rising temperature, indicating enhanced mobility of grain boundary migration. This thermal activation synergizes with confinement effects that elevate the melting point of Bi up to 900 K under vdW pressure (as illustrated in Fig. S7). We therefore propose vdW squeezing at high temperature (800-900 K) to leverage high diffusion rates and promote rapid grain boundary elimination, facilitating single-crystal perfection.

*Conclusions*—In summary, we employed MD simulations to uncover the atomic-scale mechanisms of anisotropic crystallization and phase transitions under vdW squeezing. Our simulations reproduced the pressure-dependent thickness modulation and predicted two critical phase transitions: an α-to-β transformation at 1.64 GPa and a subsequent collapse into a single-atomic-layer phase at 2.19 GPa. The anisotropic crystallization proceeds via sequential ordering—out-of-plane layering driven by confinement, followed by in-plane ordering governed by directional enthalpic interactions. Simulations involving $MoS_2$ substrates revealed that specific epitaxial angles (15° and 45°) are energetically favored, enabling single-crystal formation via orientational selection. Elevated temperatures, supported by vdW-induced melting point elevation, enhance grain boundary migration and defect healing. These findings provide atomistic insight into how confinement, substrate interactions, and thermal activation synergistically enable high-quality non-vdW 2D crystal synthesis.

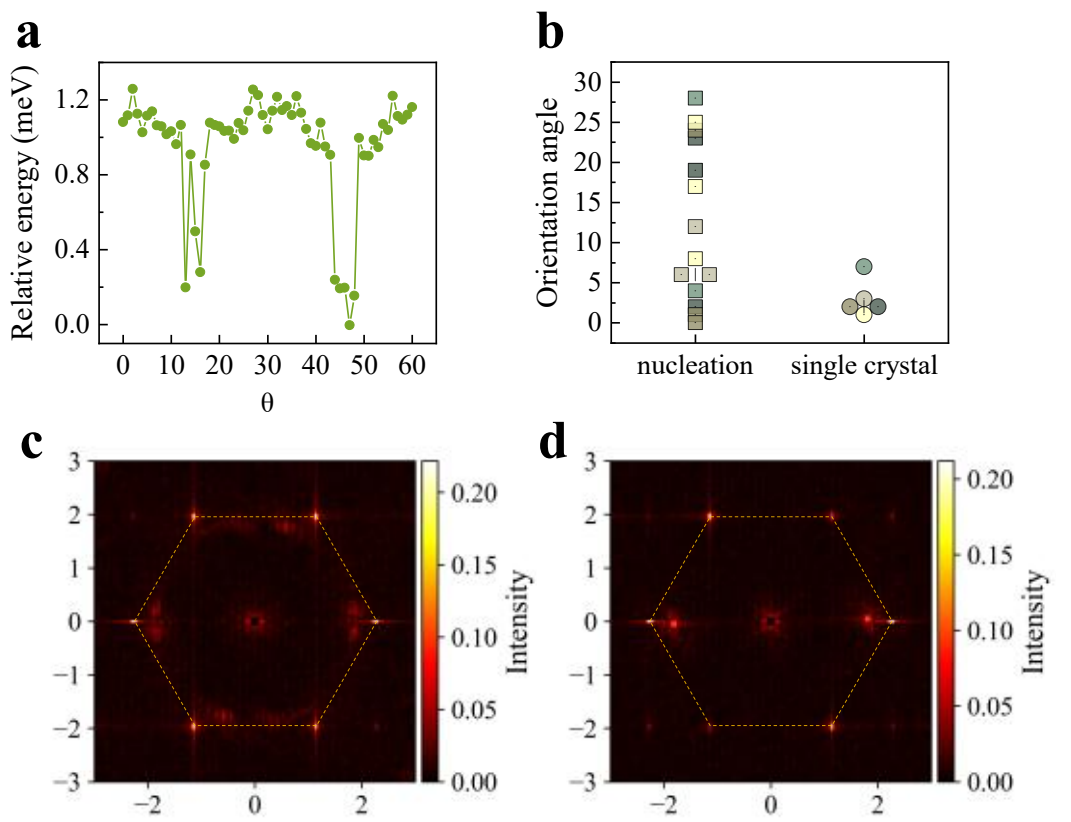

FIG. 4. Orientational selection in metal growth induced by substrate-metal interaction. (a) Total energy of the $MoS_2$/Bi/$MoS_2$ heterostructure as a function of θ, which is defined as the twist angle between the Bi [100] direction and the $MoS_2$ zigzag axis. (b) The orientation angles between the Bi lattice and the $MoS_2$ layers at both the initial nucleation stage and final single crystal perfection stage. (c) and (d) the simulated diffraction patterns for initial nucleation and final single crystal stages, respectively.

*Acknowledgments*—This work was supported by the National Key R&D Program of China (Grants No. 2021YFA0718900), National Nature Science Foundation of China (Grants No. 12374096 and No. 92477114). We also thank DP Technology for supplying computational resources via Bohrium online platform.

*Contact author: zczhong@ustc.edu.cn

*Contact author: zczhong@ustc.edu.cn